\documentclass[twocolumn,prb]{revtex4-2}
\usepackage{amsfonts}
\usepackage[T1]{fontenc}
\usepackage{amsmath,amsbsy,amssymb,graphicx}
\usepackage{times}

\begin{document}

\title{Nonlinear topological Toda quasicrystal}
\author{Motohiko Ezawa}
\affiliation{Department of Applied Physics, University of Tokyo, Hongo 7-3-1, 113-8656,
Japan}

\begin{abstract}
Topological edge states are known to emerge in certain quasicrystals. We
investigate a topological quasicrystal in the presence of nonlinearity by
generalizing the Toda lattice to include modulated periodic hoppings, where
the period is taken irrational to the original lattice. It is found that
topological edge states in a quasicrystal survive against nonlinearity based
on the quench dynamics. It is also found that an extended-localization
transition is induced by the quasicrystal hopping modulation. The present
model is experimentally realizable by a transmission line with variable
capacitance diodes, where the inductance is modulated.
\end{abstract}

\maketitle

\section{Introduction}

Quasicrystal is an ordered crystal which has no periodicity. Quasicrystals
can be opological\cite%
{Kraus,KrausT,Lang,Verbin,Madsen,Sagi,Dare,Bandres,Longhi}. One of the
simplest models is the Aubry-Andr\'{e}-Harper (AAH) model\cite{Harper,AA},
which is a one-dimensional (1D) lattice model with a periodic on-site
potential whose period is irrational to the original lattice. A 1D
quasicrystal is understood in terms of a corresponding 2D ancestor model.
For example, the AAH model has the corresponding Harper model\cite{Harper}
as an ancestor, which is a Chern insulator. In a similar way, the
off-diagonal AAH model\cite{Kraus,Verbin,Lang} has a corresponding Chern
insulator as an ancestor. They are characterized by the Hofstadter diagram
and the emergence of chiral edge states. Localization occurs in these
strongly modulated quasicrystals.

Nonlinear topological physics is an emerging field of topological physics.
It is realized in photonic\cite%
{Ley,Zhou,MacZ,HadadB,Smi,Tulo,Kruk,NLPhoto,Kirch,TopoLaser}, mechanical\cite%
{Snee,PWLo,MechaRot,Sin} and electric circuit\cite{Hadad,Sone,TopoToda}
systems. Although the Toda lattice\cite{Toda,Toda2,Toda3} is a
nontopological nonlinear system, there is a rich variety of generalizations
by introducing inhomogeneous hoppings. For instance, a topological Toda
lattice can be constructed\cite{TopoToda} by dimerizing the Toda lattice as
in the case of the Su-Schrieffer-Heeger model.

Quench dynamics starting from a localized state at an edge site presents a
strong signal to detect whether it is topological or trivial even in
nonlinear systems. The state almost remains at the initial edge site in the
topological phase because of the existence of the localized topological edge
states. On the other hand, the state rapidly spreads into the bulk in the
trivial phase. This method has proved to be useful in photonic\cite%
{NLPhoto,TopoLaser}, mechanical\cite{MechaRot,Sin} and electric circuit\cite%
{TopoToda} systems.

In this paper, we study an effect of nonlinearity in a topological
quasicrystal. We introduce a hopping modulation to the Toda lattice, which
makes it a nonlinear quasicrystal when the period is irrational to the
original lattice. In the linear limit, this model is reduced to the
off-diagonal AAH model, which has topological edge states. We study the
quench dynamics starting from an edge or bulk site. We find that there are
distinct behaviors between them. There remains a finite oscillation at the
initial site for the quench dynamics starting from the edge site. However,
right and left going propagating waves are dominant for the quench dynamics
starting from the bulk site. It evidences the emergence of the topological
edge states even in nonlinear systems. The present model is experimentally
realizable in an electric circuit with the use of the variable capacitance
diodes as in the case of the original Toda model\cite%
{Hirota,Nakajima,Singer,Yemele,Yemele2,Pelap05,Houwe}.

\section{Generalized Toda lattice}

We propose the generalized Toda model in the form of%
\begin{equation}
\frac{1}{\xi }\frac{d^{2}}{d\tau ^{2}}\left( \log \left[ 1+\xi \psi _{n}%
\right] \right) =[M_{nm}-\sum_{n}M_{nm}]\psi _{m},  \label{TodaA}
\end{equation}%
where $M_{nm}$ represents a hopping from the lattice site $m$ to $n$.
Nonlinearity is controlled by the parameter $\xi $, where the linear limit
is given by $\xi \rightarrow 0$. When we choose the hopping matrix as%
\begin{equation}
M=\kappa \sum_{n}\left( \left\vert n\right\rangle \left\langle
n+1\right\vert +\left\vert n+1\right\rangle \left\langle n\right\vert
\right) ,  \label{OriToda}
\end{equation}%
the original Toda model is recovered. All quantities are assumed to be
dimensionless.

We investigate the quench dynamics by imposing the initial condition,%
\begin{equation}
\psi _{n}\left( 0\right) =\delta _{n,n_{0}},\quad \dot{\psi}_{n}\left(
0\right) =0.  \label{IniConX}
\end{equation}%
Namely, we explore the time evolution of an input starting from the initial
site $n_{0}$ at the initial time $\tau =0$.

\section{Off-diagonal Aubry-Andr\'{e}-Harper Model}

In the present work, we study the generalized Toda model (\ref{TodaA}),
where the hopping matrix is given by the off-diagonal AAH model\cite%
{Harper,AA,Kraus,Verbin},%
\begin{equation}
M_{\phi }=\sum_{n}J_{n}^{\phi }\left( \left\vert n\right\rangle \left\langle
n+1\right\vert +\left\vert n+1\right\rangle \left\langle n\right\vert
\right) ,  \label{AAH}
\end{equation}%
with%
\begin{equation}
J_{n}^{\phi }=\kappa +J\cos \left( 2\pi \alpha n+\phi \right) ,  \label{Jn}
\end{equation}%
where the hopping amplitude $J_{n}^{\phi }$ is modulated with the period $%
1/\alpha $ with $0<\alpha <1$, $0\leq J/\kappa \leq 1$ and $\phi $ is the
phase of the modulation. It is reduced to the original Toda model\cite%
{Toda,Toda2,Toda3} when $J=0$, and to the dimerized Toda model\cite{TopoToda}
when $\alpha =1/2$. The original Toda system is not topological, while the
dimerized Toda system is topological\cite{TopoToda}.

The off-diagonal AAH model is equivalent to the 2D ancestor model given by%
\cite{KrausT}%
\begin{eqnarray}
H^{\text{2D}} &=&\sum_{n,m}(\kappa \left\vert n,m\right\rangle \left\langle
n+1,m\right\vert  \notag \\
&&+\frac{J}{2}e^{2\pi i\alpha n}\left\vert n,m\right\rangle \left\langle
n+1,m+1\right\vert  \notag \\
&&+\frac{J}{2}e^{-2\pi i\alpha n}\left\vert n,m\right\rangle \left\langle
n+1,m-1\right\vert )+\text{h.c.},  \label{2D-model}
\end{eqnarray}%
where the modulation phase $\phi $ has been traded with an extra synthetic
dimension. We can confirm the equivalence by introducing the Fourier
transformation%
\begin{equation}
\left\vert m\right\rangle =\sum_{\phi }e^{-im\phi }\left\vert \phi
\right\rangle
\end{equation}%
into Eq.(\ref{2D-model}), and rewrite it as%
\begin{equation}
H^{\text{2D}}=\sum_{\phi }M_{\phi }\left\vert \phi \right\rangle
\left\langle \phi \right\vert ,
\end{equation}%
where $M_{\phi }$ is given by Eq.(\ref{AAH}). The model (\ref{2D-model})
describes a Chern insulator\cite{KrausT} for any $\alpha $, and hence, the
off-diagonal AAH model (\ref{AAH}) also describes a topological insulator.
Note that the edge states in the off-diagonal AAH model are mapped to the
chiral edge states in the 2D ancestor model.

We confirm this correspondence numerically by considering a finite chain
with length $L$ in the rest of this section.

\begin{figure}[t]
\centerline{\includegraphics[width=0.44\textwidth]{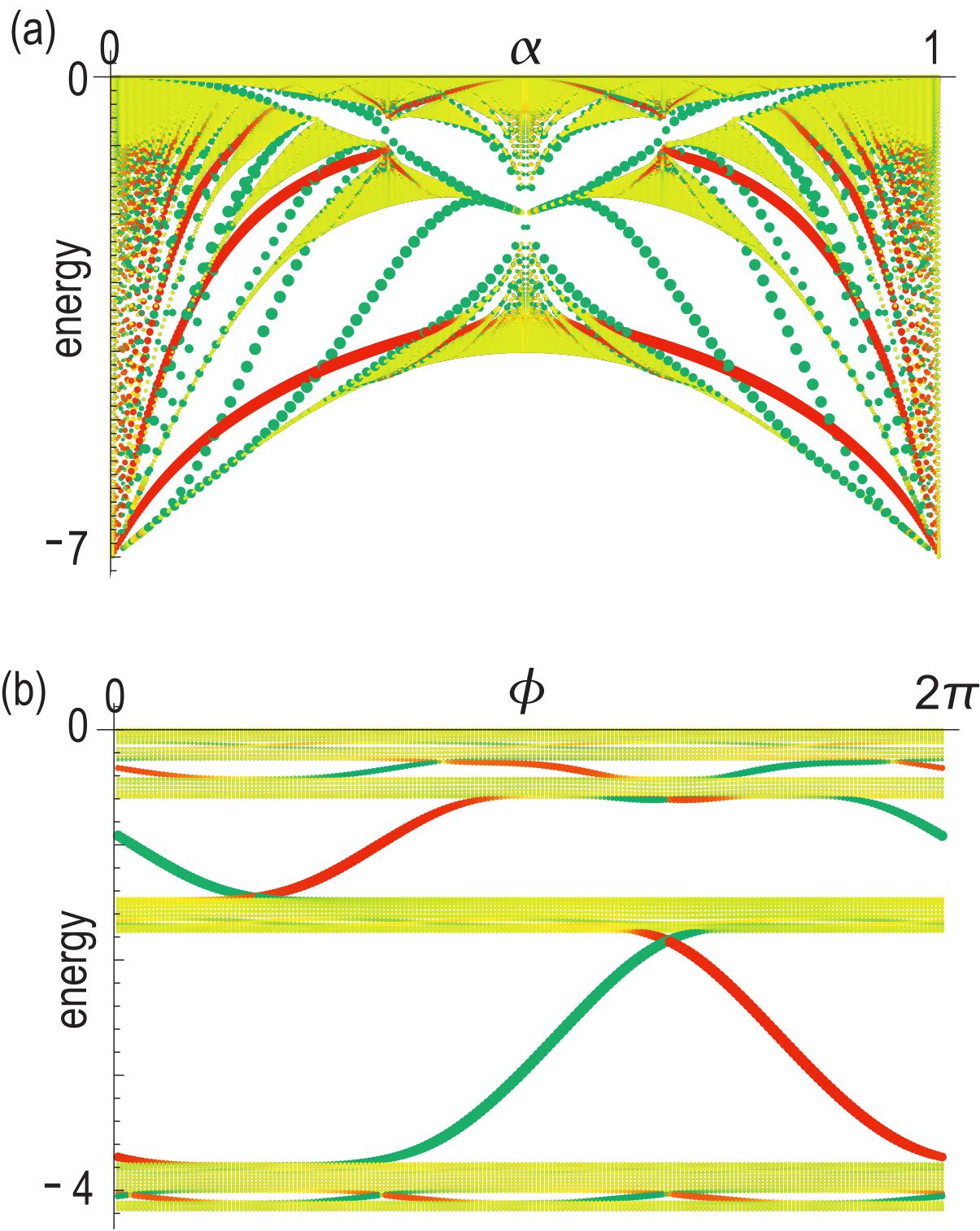}}
\caption{(a) Energy spectrum of the off-diagonal AAH model as a function of
the phase $\protect\alpha $ at $\protect\phi =0$. It presents a Hofstadter
butterfly. (b) Energy spectrum as a function of $\protect\phi $ at $\protect%
\alpha =(\protect\sqrt{5}-1)/2$. The energy is in units of $\protect\kappa $%
. Curves in red (green) indicate the left(right)-localized edge states,
while those in lime green indicate the bulk states. We have taken a finite
chain with length $L=100$, and set $J=0.75\protect\kappa$. }
\label{FigHof}
\end{figure}

\begin{figure*}[t]
\centerline{\includegraphics[width=0.98\textwidth]{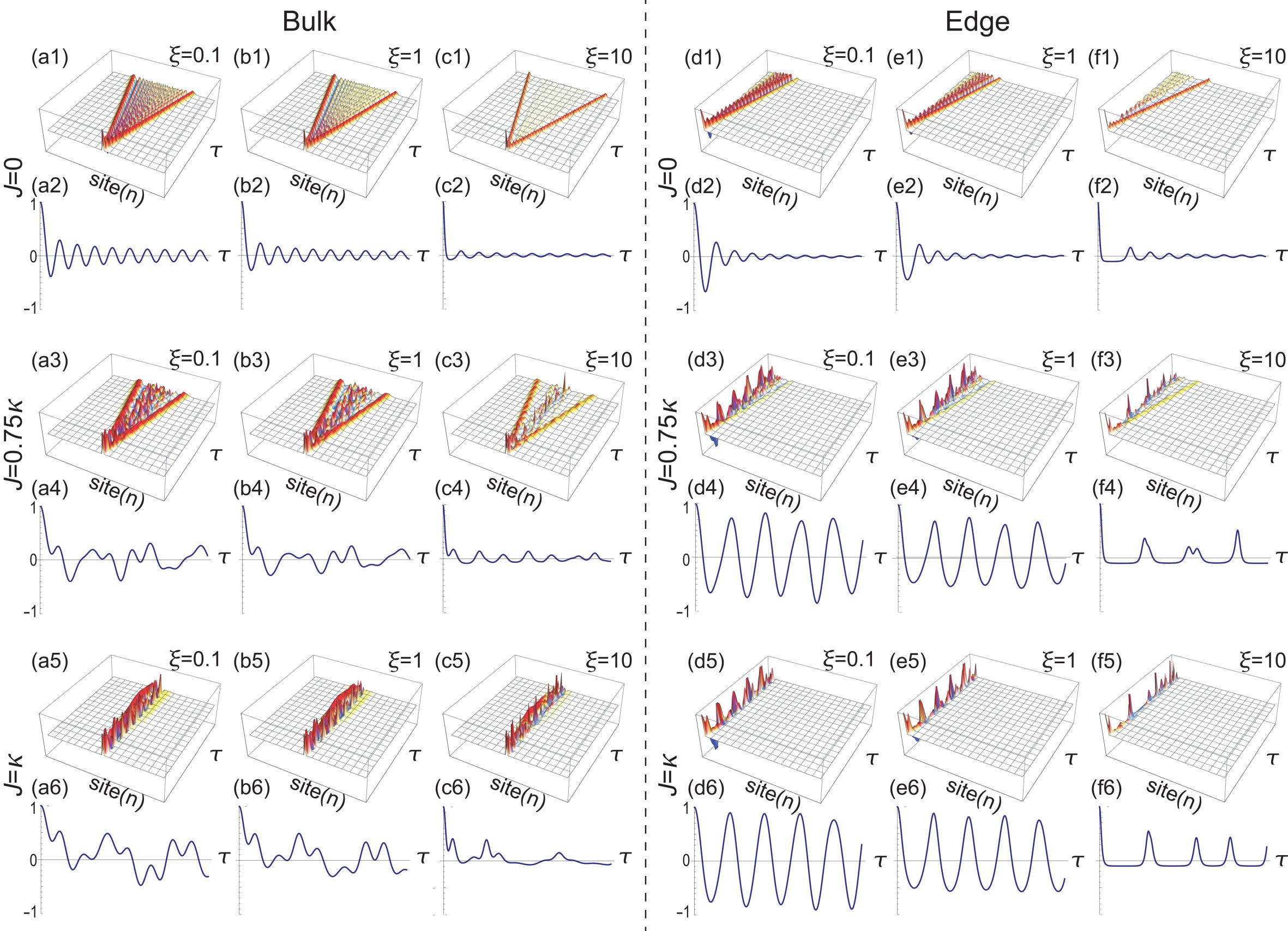}}
\caption{ Time evolution of $\protect\psi _{n}$ starting from the edge site $%
n_{0}=1$ for the left panel, and from the bulk site $n_{0}=L/2$ for the
right panel. (a1)$\sim $(f1), (a3)$\sim $(f3) and (a5)$\sim $(f5) Spatial
profile of the time evolution of $\protect\psi _{n}$. (a2)$\sim $(f2), (a4)$%
\sim $(f4) and (a6)$\sim $(f6) The time evolution of $\protect\psi _{n_{0}}$
at the initial site. We have taken a finite chain with length $L=100$. We
have set $\protect\phi =0$ and $T=30$. (a1)$\sim $(f2) $J=0$, (a3)$\sim $%
(f4) $J=0.75\protect\kappa $ and (a5)$\sim $(f6) $J=\protect\kappa $. There
is no localization phenomenon nor edge enhancement in the original Toda
model ($J=0$). The localization phenomenon occurs due to the quasicrystal
hopping modulation for $J\neq 0$\ but it is small for small $J$. It reaches
the maximum at $J=\protect\kappa $. An extended-localization transition
point is at $J=0.76\protect\kappa $, about which we refer to Fig.\protect\ref%
{FigMaxJ}(d). The edge enhancement occurs for all $J\neq 0$\ due to the
emergence of topological edge states.}
\label{FigPropagation}
\end{figure*}

The eigenspectrum of the matrix $M$ in Eq.(\ref{AAH}) is calculated and
shown in Fig.\ref{FigHof}(a) as a function of rational number $\alpha $. The
result is a well-known Hofstadter diagram, consisting of edge states marked
in red and green in addition to the bulk states marked in lime green. The
Hofstadter diagram is peculiar to a Chern insulator in two dimensions.

The system describes a quasicrystal when $\alpha $ is taken to be an
irrational number. For definiteness, we explicitly take $\alpha $ to be the
inverse golden ratio $\left( \sqrt{5}-1\right) /2$. The energy spectrum is
shown as a function of $\phi $ in Fig.\ref{FigHof}(b). We find
left-localized and right-localized edge states as a function of $\phi $, as
depicted by red and green curves connecting two bulk bands. Note that these
edge states are present for all $\phi $. This energy spectrum is also
peculiar to a Chern insulator with the left-moving and right-moving chiral
edge states in two dimensions.

The linear limit ($\xi =0$) of Eq.(\ref{TodaA}) with (\ref{AAH}) is the
dynamical AAH model, where the above topological property holds as it is. In
what follows, we explore how these topological edge states survive as the
nonlinearity $\xi $ increases.

\section{Quench dynamics}

We first investigate the quench dynamics in the original Toda lattice, where 
$J=0$. The time evolution of an input starting from the bulk site $n_{0}=L/2$
is shown in Fig.\ref{FigPropagation}(a1)$\sim$(c1). There are only
propagating waves and no localized modes. In weak nonlinearity ($\xi =0.1$),
there are ripples between the two wave fronts as in Fig.\ref{FigPropagation}%
(a1). In strong nonlinearity ($\xi =10$), on the other hand, there are no
ripples because solitary waves are formed as in Fig.\ref{FigPropagation}%
(c1). We also show the time evolution at the initial site in Fig.\ref%
{FigPropagation}(a2)$\sim$(c2). It decreases rapidly with oscillations.

The time evolution starting from the edge site $n_{0}=1$ is shown in Fig.\ref%
{FigPropagation}(d1)$\sim $(f1). The time evolution at the initial site is
similar between the quench dynamics starting from the bulk and edge sites as
shown in Fig.\ref{FigPropagation}(a2)$\sim $(f2),\ which indicates that
there are no edge states with $J=0$ in consistent with the fact that the
system is not topological.

Next, we study the quench dynamics starting from the bulk site in the case
of $J\neq 0$, where the system is topological. There appear right and left
going propagating waves in addition to localized modes in the vicinity of
the initial site, as shown in Fig.\ref{FigPropagation}(a3)$\sim $(c3). The
localized mode emerges due to the effect of the quasicrystal hopping
modulations. In addition, by comparing Fig.\ref{FigPropagation}(a1)$\sim $%
(c1) and Fig.\ref{FigPropagation}(a3)$\sim $(c3), the velocity of the wave
propagation becomes smaller in the presence of the quasicrystal modulation.
It may be also due to the interference of the hopping modulation.

The time evolution at the edge is shown in Fig.\ref{FigPropagation}(d3)$\sim 
$(f3). The quench dynamics starting from the edge site is quite different
from that starting from the bulk site. In contrast to the quench dynamics
starting from the bulk site shown in Fig.\ref{FigPropagation}(a3)$\sim $%
(c3), the intensity of the propagating wave is weak but almost all states
are localized at the edge as shown in Fig.\ref{FigPropagation}(d3)$\sim $%
(f3). The amplitude of oscillation is stationary after a certain time as in
Fig.\ref{FigPropagation}(d4)$\sim $(f4), which is larger than that of the
bulk shown in Fig.\ref{FigPropagation}(a4)$\sim $(c4). These phenomena
indicate the emergence of the topological edge states even in a nonlinear
quasicrystal.

Then, we study the quench dynamics in the case of $J=\kappa $. The state is
almost localized at the initial site for the quench dynamics starting from
both the edge and bulk sites, as shown in Fig.\ref{FigPropagation}(a5)$\sim $%
(f5). Especially, amplitude of the oscillation is quite large as shown in
Fig.\ref{FigPropagation}(d6)$\sim $(f6). It indicates that the strong
localization occurs due to the strong quasicrystal hopping modulation.

We proceed to study systematically the effects due the quasicrystal hopping
modulation and the nonlinearity by introducing an index%
\begin{equation}
\Psi \equiv \max_{T/2<t<T}(\left\vert \psi _{n_{0}}\right\vert ),
\end{equation}%
which is the maximum value of $|\psi _{n_{0}}|$ in the time span $T/2<\tau
<T $, where $T$ is large enough so that the oscillation is stationary.

\begin{figure}[t]
\centerline{\includegraphics[width=0.48\textwidth]{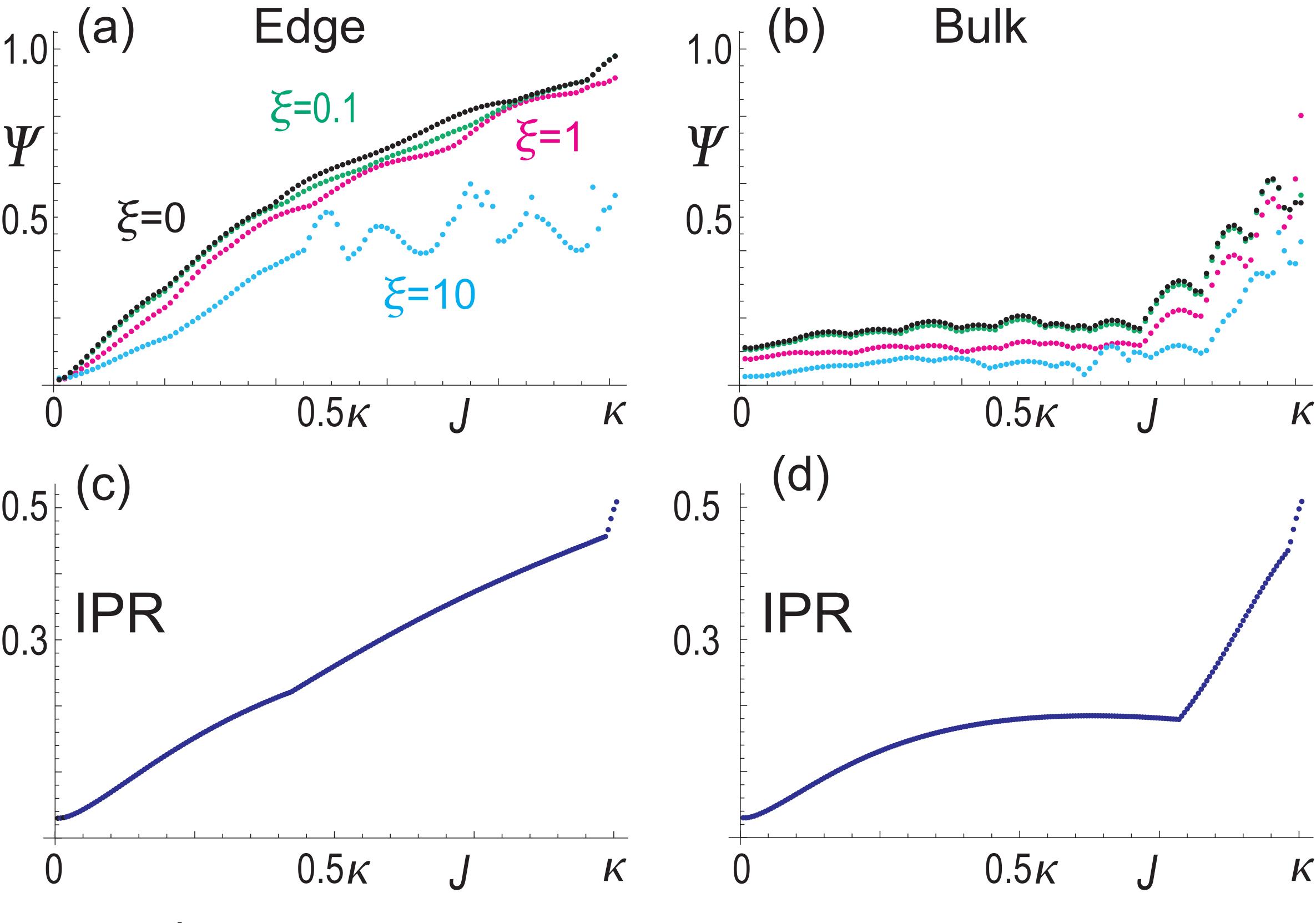}}
\caption{Index $\Psi $ as a function of $J$ for various $\protect\xi $ at
(a) the edge site $n_{0}=1$ and (b) bulk site $n_{0}=L/2$ . Black curve
indicates $\protect\xi =0$, green curve indicates $\protect\xi =0.1$,
magenta curve indicates $\protect\xi =1$ and cyan curve indicates $\protect%
\xi =10$. We have set $T=50$ and $L=100$. The horizontal axis is $J$ and the
vertical axis is $\Psi $. (c) The IPR as function of $J$ for $P$, and (d)
that for $P_{\text{bulk}}$.}
\label{FigMaxJ}
\end{figure}

\begin{figure}[t]
\centerline{\includegraphics[width=0.48\textwidth]{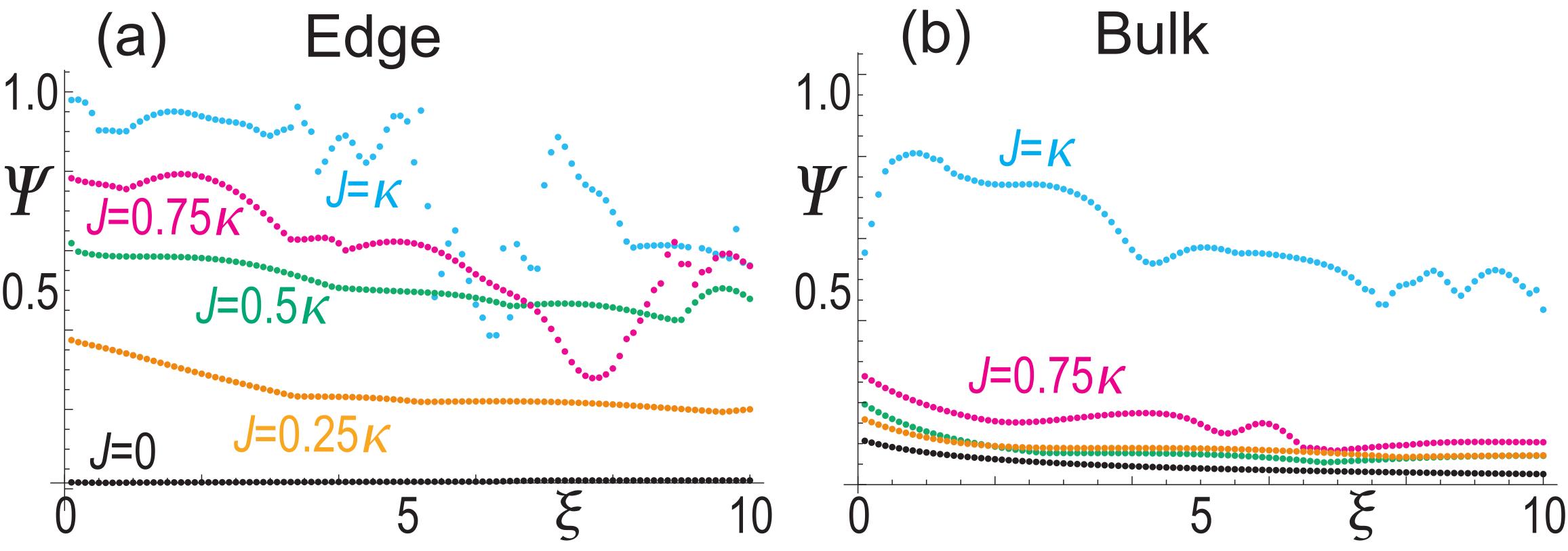}}
\caption{Index $\Psi $ as a function of $\protect\xi $ for various $J$ at
(a) the edge site $n_{0}=1$ and (b) bulk site $n_{0}=L/2$. Black curve
indicates $J=0$, orange curve indicates $J=0.25\protect\kappa $, green curve
indicates $J=0.5\protect\kappa $, magenta curve indicates $J=0.75\protect%
\kappa $ and cyan curve indicates $J=\protect\kappa $. We have set $T=50$
and $L=100$. The horizontal axis is $\protect\xi $ and the vertical axis is $%
\Psi $. }
\label{FigMaxZ}
\end{figure}

The index $\Psi $ is shown as a function of the periodic-modulation
magnitude $J$ for various nonlinearity $\xi $ in Fig.\ref{FigMaxJ}(a) and
(b), where the quench dynamics starting from the edge and bulk sites are
studied. There are some distinguishable features between them. First, $\Psi $
is larger for the quench dynamics starting from the edge site. It indicates
the existence of the topological edge states in the nonlinear quasicrystal.
Next, $\Psi $ increases as a function of $J$ in both the cases, which
indicates that the localization is enhanced in the presence of the
quasiperiodic potential.

In order to understand the localization phenomenon\ in the quasicrystal, we
calculate the inverse-participation ratio (IPR)\ defined by%
\begin{equation}
P\equiv \sum_{n=1}^{L}\left\vert \tilde{\psi}_{n}\right\vert ^{4},
\label{IPR}
\end{equation}%
where $\tilde{\psi}_{n}$ is the normalized eigenstate of $M_{\phi }$. The IPR
satisfies $0\leq P\leq 1$. The IPR measures the magnitude of the
localization, where the system is localized when $P$ is close to 1, while it
is extended when $P$ is close to 0. We show the IPR as a function of $J$ in
Fig.\ref{FigMaxJ}(c). It linearly increases as a function of $J$, which is
consistent with Fig.\ref{FigMaxJ}(a).

The IPR defined by Eq.(\ref{IPR}) contains a large contribution from the
edge, which is irrelevant to the localization phenomenon. Hence, we define
the IPR for the bulk by%
\begin{equation}
P_{\text{bulk}}\equiv \sum_{n=2}^{L-1}\left\vert \tilde{\psi}_{n}\right\vert
^{4},
\end{equation}%
which is shown in Fig.\ref{FigMaxJ}(d). It exhibits a clear transition
around $J\simeq 0.76\kappa $. It is an extended-localization transition
induced by the quasicrystal hopping modulation $J_{n}^{\phi }$. The quench
dynamics at $J=0.75\kappa $\ in Fig.(a3)$\sim $(f3) is the one near this
transition point.

Finally, in order to study the effect of the nonlinearity on the
localization due to the quasicrystal hopping modulation $J_{n}^{\phi }$, we
show $\Psi $ as a function of $\xi $ for various $J$ in Fig.\ref{FigMaxZ}. $%
\Psi $ remains almost unchanged even when nonlinearity increases, which
indicates that the topological edge states survive for strong nonlinearity.

\section{Electric-circuit realization}

The original Toda lattice is realized by a transmission line with the use of
variable-capacitance diodes and inductors\cite{NLAnderson,TopoToda}. Here we
show how to realize the generalized Toda lattice with the use of
variable-capacitance diodes and inductors.

\begin{figure}[t]
\centerline{\includegraphics[width=0.48\textwidth]{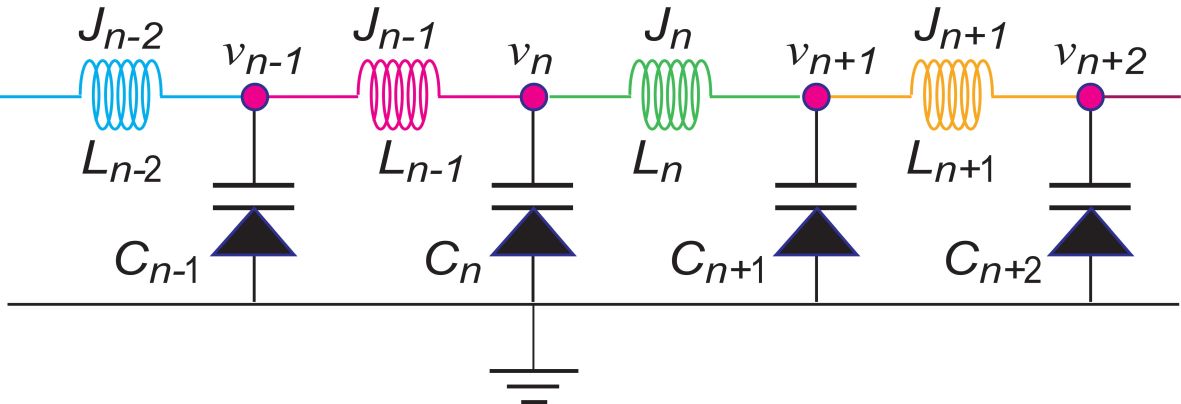}}
\caption{Illustration of a transmission line made of nonlinear elements
realizing the generalized Toda lattice. The inductance is modulated, as
indicated by colors. Each node is grounded via a variable-capacitance diode.}
\label{FigIllust}
\end{figure}

We consider a transmission line as shown in Fig.\ref{FigIllust}. The
Kirchhoff law is given by 
\begin{align}
L_{n}\frac{dJ_{n}}{dt}=& v_{n}-v_{n+1}, \\
\frac{dQ_{n}}{dt}=& J_{n-1}-J_{n},
\end{align}%
where $v_{n}$ is the voltage, $J_{n}$ is the current and $Q_{n}$ is the
charge at the node $n$, while $L_{n}$ is the inductance for the inductor
between the nodes $n$ and $n+1$, as illustrated. The Kirchhoff law is
summarized in the form of the second-order differential equation\cite%
{NLAnderson,TopoToda}, 
\begin{align}
\frac{d^{2}Q_{n}}{dt^{2}}=& \frac{dJ_{n-1}}{dt}-\frac{dJ_{n}}{dt}  \notag \\
=& \frac{1}{L_{n-1}}\left( V_{n-1}-V_{n}\right) -\frac{1}{L_{n}}\left(
V_{n}-V_{n+1}\right) ,  \label{EqC}
\end{align}%
where we have introduced a new variable $V_{n}$ by $v_{n}=V_{0}+V_{n}$.

The capacitance is a function of the voltage $V_{n}$\ in the
variable-capacitance diode, and it is well given by\cite{Nakajima}%
\begin{equation}
C\left( V_{n}\right) =\frac{Q\left( V_{0}\right) }{F_{0}+V_{n}-V_{0}},
\label{CV}
\end{equation}%
where $F_{0}$ is a constant characteristic to the variable-capacitance
diode. Especially, we have%
\begin{equation}
C\left( V_{0}\right) =\frac{Q\left( V_{0}\right) }{F_{0}}.
\end{equation}%
The charge is given by%
\begin{equation}
Q_{n}=\int_{0}^{V_{n}}C\left( V\right) dV=Q\left( V_{0}\right) \log \left[ 1+%
\frac{V_{n}}{F_{0}}\right] +\text{const.}  \label{EqD}
\end{equation}%
It follows from Eqs.(\ref{EqC}) and (\ref{EqD}) that 
\begin{align}
Q& \left( V_{0}\right) \frac{d^{2}}{dt^{2}}\log \left[ 1+\frac{V_{n}}{F_{0}}%
\right]  \notag \\
& =\frac{V_{n+1}}{L_{n}}-\left( \frac{1}{L_{n-1}}+\frac{1}{L_{n}}\right)
V_{n}+\frac{V_{n-1}}{L_{n-1}}.  \label{EqA}
\end{align}%
This is a closed form of the differential equation for $V_{n}$.

Eq.(\ref{EqA}) has the same form as the generalized Toda equation (\ref%
{TodaA}) with%
\begin{equation}
M_{nm}=\sum_{n}\kappa _{n}\left( \left\vert n\right\rangle \left\langle
n+1\right\vert +\left\vert n+1\right\rangle \left\langle n\right\vert
\right) ,
\end{equation}%
when we set 
\begin{eqnarray}
\tau &=&t/\sqrt{L_{0}Q\left( V_{0}\right) V_{1}^{2}/F_{0}},\quad \psi
_{n}=V_{n}/V_{1},  \notag \\
\xi &=&V_{1}/F_{0},\qquad \kappa _{n}=L_{0}/L_{n}.
\end{eqnarray}%
where $1/L_{0}$ is the mean of $1/L_{n}$. For instance, by choosing the
inductance as%
\begin{equation}
1/L_{n}=\kappa +J\cos \left( 2\pi \alpha n+\phi \right) ,
\end{equation}%
Eq.(\ref{Jn}) is reproduced.

Finally, the initial condition (\ref{IniConX}) is rewritten as%
\begin{equation}
V_{n}=V_{1}\delta _{n,1}\quad \text{and}\quad \dot{V}_{n}=0\quad \text{at}%
\quad \tau =0.
\end{equation}%
Hence, the nonlinearity $\xi $\ is controlled only by changing the initial
voltage $V_{1}$ without changing samples in experiments. The quench dynamics
is experimentally observed by measuring the time evolution of the voltage.

\section{Conclusion}

We have investigated topological properties of a nonlinear quasicrystal by
generalizing Toda lattice. The existence of the topological edge states is
well signaled by comparing the quench dynamics starting from an edge site
and a bulk site. Stationary oscillation occurs at the edge even in nonlinear
regime, which indicates that the system is a nonlinear topological
quasicrystal. The present model is experimentally realizable based on
electric circuits with variable-capacitance diodes and inductors. It is
interesting to study further interplay among, topology, quasicrystal and
nonlinearity.

The author is very much grateful to M. Kawamura, S. Katsumoto and N. Nagaosa
for helpful discussions on the subject. This work is supported by the
Grants-in-Aid for Scientific Research from MEXT KAKENHI (Grants No.
JP17K05490 and No. JP18H03676). This work is also supported by CREST, JST
(JPMJCR16F1 and JPMJCR20T2).

\end{document}